\documentclass[global,twocolumn]{svjour}
\usepackage{graphicx}
\usepackage{amsmath}
\usepackage{amssymb}

\newcommand{\eref}[1]{(\ref{#1})}

\newcommand{\nnn}{\nonumber\\}
\newcommand{\PO}{\;.}   
\newcommand{\CO}{\;,}   

\newcommand{\oh}{{\frac{1}{2}}} %

\newcommand{\rme}{\mathrm{e}}
\newcommand{\rmi}{\mathrm{i}}
\newcommand{\rmd}{\mathrm{d}}

\newcommand{\Or}{\mathord{\mathcal{O}}}

\newcommand{\sinc}{\mathop{\mathrm{sinc}}\nolimits}

\newcommand{\Gammafunction}{\mathop{\mathrm{\Gamma}}\nolimits}

\newcommand{\Natural}{{{\mathbb{N}}}}
\newcommand{\Z}{{{\mathbb{Z}}}}

\newcommand{\kB}{{k}_{\rm B}}
\newcommand{\Angstrom}{{\mathrm{\AA}}}
\newcommand{\LT}{L_\lambda}
\newcommand{\vm}{v_{\rm m}}
\newcommand{\vg}{v_{\rm g}}
\newcommand{\vgt}{\tilde{v}_{\rm g}}
\newcommand{\mg}{m_{\rm g}}
\newcommand{\pv}{p_{0}}
\newcommand{\Deltar}{\delta r}

\def\makeheadbox{{%
\hbox to0pt{\vbox{\baselineskip=10dd\hrule\hbox
to\hsize{\vrule\kern3pt\vbox{\kern3pt
\hbox{Preprint -- July 31, 2003}
\hbox{}
\kern3pt}\hfil\kern3pt\vrule}\hrule}%
\hss}}}

\begin{document}
\sloppy

\title{Decoherence in a Talbot Lau interferometer: the influence of molecular scattering}

\author{
Lucia Hackerm\"uller \and Klaus Hornberger \and Bj\"orn Brezger\thanks{present address: Fachbereich Physik, Universit{\"a}t
Konstanz, D-78457 Konstanz } \and Anton Zeilinger \and and Markus
Arndt
}

\institute{Institut f\"ur Experimentalphysik der
Universit\"at Wien, Boltzmanngasse 5, A-1090 Wien}

\date{\ }

\maketitle

\begin{abstract}
We study the interference of C$_{70}$ fullerenes in a Talbot-Lau
interferometer with a large separation between the diffraction
gratings. This permits the observation of recurrences of the
interference contrast both as a function of the de Broglie
wavelength and in dependence of the interaction with background
gases. We observe an exponential decrease of the fringe
visibility with increasing background pressure and find good
quantitative agreement with the predictions of decoherence
theory. From this we extrapolate the limits of matter wave
interferometry and conclude that the influence of collisional
decoherence may be well under control in future experiments with
proteins and even larger objects.
\end{abstract}

\section{Introduction}
\label{sec:intro} Matter wave interferometry of small quantum
objects has become an active field of research during the last
decades~\cite{Berman1997a}. The new field of coherent optics with
large molecules is now exploring the technical and possibly
fundamental limits of interferometry with quantum objects of high
mass, high internal complexity and high internal excitation --
i.e. with novel properties which allow to study in detail the
quantum-classical transition~\cite{Arndt1999a,Arndt2002a}.

Several challenges arise when one moves from small to large
systems: At a given velocity the de Broglie wavelength
$\lambda=h/(m v)$ shrinks with increasing mass, and in addition
it becomes increasingly difficult to slow down massive systems
which have a large kinetic energy. The requirements on
interferometer technology for de Broglie wavelengths in the
sub-picometer range are already rather demanding. But even more
important are the potential mechanisms which may lead to
decoherence, that is to a loss of visibility in the interference
pattern due to the coupling of the quantum system to its
environment (see e.g. \cite{Joos1985a,Zurek1991a}).

The investigation of this apparent loss of quantum properties has
become an important corner stone of modern quantum physics -- not
only due to its fundamental role in mesoscopic
physics and its importance for the understanding of the
quantum-classical transition, but also because of its potential
impact on emerging quantum technologies, such as quantum
computers.

Any increase in size and complexity generally opens new
decoherence channels, and for large molecules one can think of
many interactions with the environment, either by scattered
radiation \cite{Chapman1995b}, by collisions with particles
\cite{Hornberger2003a}, by an interaction with fluctuating
quasi-static electro-magnetic fields \cite{Myatt2000a} or even by
the interaction with gravitational waves \cite{Reynaud2002a}.

While it is impossible to manipulate and track the details of the
perturbations for really macroscopic systems, the environment of
isolated mesoscopic quantum systems can still be efficiently
controlled. In the present paper we focus on one particular
interaction between large molecules and an environment, namely
collisions between the coherently propagating molecules and
various background gases. First results on this subject have
already been discussed in a previous letter
\cite{Hornberger2003a}. In the present work we give more detailed
background information on our quantitative investigations of
collisional decoherence of the fullerene C$_{70}$, and we study
both experimentally and theoretically the influence of increased
interaction times, which will be unavoidable in interferometry
with proteins.

\section{The Talbot Lau interferometer}
\label{sec:1Setup}
\begin{figure}
\centering
\includegraphics[width=1.1\columnwidth]{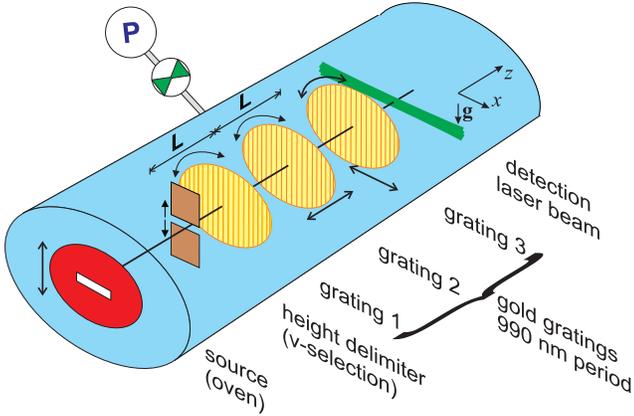}
\caption{Artist's view of the 'pressurized' Talbot Lau
interferometer. A beam of C$_{70}$ molecules is generated from
fullerene powder at 900 K. The beam passes a series of three gold
gratings each with a grating constant of $d=990\,$nm, an open
width of 480 nm and a grating thickness of 500 nm. The grating
separation $L$ was set to 38\,cm. The whole
vacuum chamber is evacuated to $2\times 10^{-8}$\,mbar and can
then be pressurized with different gases, typically up to
$10^{-6}$\,mbar. The fullerenes are detected using a laser
induced thermal ionization process \cite{Nairz2000a}. The
interferogram is scanned by shifting the third grating along the
grating vector.} \label{fig:1}
\end{figure}

One can conceive various experimental arrangements to demonstrate
the wave-nature of material particles and many interferometers
have already been built for atoms (see refs. in
\cite{Berman1997a}). Also for small molecules a number of
arrangements such as grating diffraction
\cite{Schollkopf1994a,Arndt1999a}, Ramsey-Bord\'e interferometry
\cite{Borde1994a,Lisdat2000a}, or Mach-Zehnder interferometry
\cite{Chapman1995b,Bruehl2002a} have been shown to work. However,
all these arrangements need  well collimated beams or
experimentally distinguishable internal states in order to
separate the various diffraction orders. This requirement makes
them less suitable for large clusters and large molecules for
which brilliant sources and highly efficient detection schemes
still have to be developed.

A near-field interferometer of the Talbot Lau type, in contrast,
does away with the collimation requirement. Unlike the far-field
interferometers, it is more compact, rugged and allows a much
higher transmission \cite{Clauser1992a}.

The basic idea of such a device, the lens-less periodic imaging
of molecular density distributions, can already be seen from a
short discussion of the Talbot effect as used in light optics
\cite{Patorski1989a}. Suppose that a plane wave $\psi_0=\exp(\rmi
k z)$ illuminates a grating located in the $(x,y)$-plane with
grating function $t(x)$. The wave function at a distance $L$
behind the grating is then given in paraxial approximation by the
Kirchhoff-Fresnel integral
\[
\psi_L = \rme^{\rmi k L} \left(\frac{k}{2\pi\rmi L}\right)^{\oh}
\int  \rmd x' t(x')
\exp\left(\rmi k\frac{(x-x')^2}{2L}\right) \PO
\]
It is easily evaluated if the grating function is periodic,
\begin{equation}
\label{eq:gratingfunction} t(x) = \sum_{\ell\in\Z} a_\ell \exp
\left( 2\pi\rmi \ell\frac{x}{d} \right) \CO
\end{equation}
and one finds by Gaussian integration,
\[
\psi_L =  \rme^{\rmi k L} \sum_\ell a_\ell \exp \left( 2\pi\rmi
\ell\frac{x}{d} \right) \exp \left( -\rmi\pi \ell^2
\frac{L\lambda}{d^2} \right) \PO
\]
From this expression one observes immediately that at even
multiples of the distance
\[
\LT =  \frac{{d^2 }}{\lambda }
\]
the transverse part of the wave function is simply given by the
grating function \eref{eq:gratingfunction}. The grating pattern
is also repeated at {odd} multiples of the \emph{Talbot length}
$L_\lambda$, but there it is shifted along $x$ by half a grating
period $d/2$.

This lens-less Talbot imaging is a pure interference effect and
was already successfully applied to material objects
\cite{Chapman1995c,Deng1999a}. However, in the version described
so far it still requires a plane wave, i.e., a parallel input
beam. The full intensity gain of the Talbot effect is only
deployed when it is applied to uncollimated and therefore much
more intense molecular beams~\cite{Clauser1994a,Brezger2002a}.
This is realized if the single diffraction grating is replaced by
three gratings, which act -- from front to end  -- as a
multiplexing collimator, a diffraction grating and a detection
mask (for details see, e.g., \cite{Brezger2003a}). Each point in
the grating then acts as the source of an interference pattern
and, even though there is no coherence between different source
points, the independent interference patterns originating from
each of them overlap in a position-synchronized manner to form a
pattern of high fringe visibility. One may also regard the first
grating as a tool to impose some coherence on the uncollimated
molecular beam. The finite width of each opening in the first
grating induces lateral coherence at the second grating which is
of the order of 2-3 grating periods.

Our experimental setup is based on this idea, and a sketch of it
is shown in Fig.~\ref{fig:1}. C$_{70}$ molecules are sublimated
at 900 K to form an effusive beam with molecular de Broglie
wavelengths in the range from 2\,pm to 5\,pm. The beam is
essentially uncollimated in the horizontal direction, but it is
selected by three spatially separated height delimiters,
namely the oven aperture (200\,$\mu$m), a central height
delimiter (alternatively 50 or 150\,$\mu$m), and the detector
laser beam with a waist of 10\,$\mu$m. By shifting the oven
vertically one can then select a well-determined free-flight
parabola and thus a certain velocity class. This way, mean
velocities could be chosen from 90\,m/s to 220\,m/s with the full
width at half maximum of their distributions ranging from 7\% to
17\% of the mean velocity.
The gravitational method is superior to a set of rotating
slotted disks because it avoids additional vibrations and allows up to 100\,\%
throughput at the central selected velocity.

\begin{figure}[t]
  \centering
  \includegraphics[width=\columnwidth]{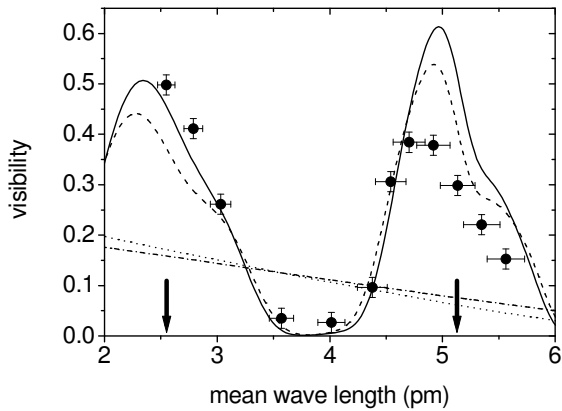}
\caption{Interferometer visibility as a function of the mean
molecular de Broglie wavelength. We observe a clear recurrence of
the interference maximum both in the experiment (circles) and in
the numerical model (lines). This is expected for the Talbot
effect with varying wavelength or varying Talbot distance
respectively. The pressure in the chamber was below $3\times
10^{-8}$\,mbar so that collisions are still negligible, and the
central height limiter was set to 50 $\mu$m. The theoretical
lines correspond to the quantum calculation at open fractions of
$f=0.45$ (solid line) and $f=0.48$ (dashed line), while the
corresponding classical expectation is shown by the dotted
($f=0.45$) and dash-dotted line ($f=0.48$). The arrows indicate
the wavelengths where the Talbot criterion $L=m \LT,=m
g^2/\lambda\;, m\in\Natural$, is met with $m=1$ for $\lambda =
2.58\,$pm  and $m=2$ for $\lambda = 5.14\,$pm. With respect to
ideal gratings the true maxima are slightly shifted to smaller
wavelengths and the two maxima have different heights. This is
due to the interaction between the molecule and the wall.}
\label{fig:2}
\end{figure}

The transverse coherence of the beam is unprepared until the
molecules pass the first grating. The three gold gratings have a
period of 990\,nm, a nominal open fraction of $0.48\pm0.02$ (as
specified by the manufacturer Heidenhain, Traunreut) and a flat
open field of roughly 16\,mm diameter. We limit the lateral width
of the molecular beam to 1\,mm which is also comparable to the
width of the ionization range of the detecting laser
beam~\cite{Nairz2000a}. The distance between consecutive gratings
was set to $L=0.38\,$m. This is  almost twice as large as in our
previous experiment~\cite{Hornberger2003a} and enables us to
observe two Talbot recurrences (see Fig.~\ref{fig:2}). All
gratings can be rotated around the molecular beam to align them
with an accuracy of about 1\,mrad both with respect to each other
and to the direction of earth's acceleration. The third grating
G$_{3}$ masks the molecular density pattern behind the second
grating. G$_{3}$ is mounted on a piezo translation stage
(Piezosystem Jena) and is scanned perpendicular to the molecular
beam in steps of 100 nm. Those molecules which are in phase with
the openings of G$_{3}$ pass the grating and are heated by the
crossing Ar$^+$ laser beam (488 nm, 15\,Watt). The positive ions
which are generated by the laser induced thermionic emission are
counted as a function of the lateral position $x_{\rm s}$ of
G$_{3}$.  As shown below, theory predicts an almost sine-wave
shaped interferogram $S(x_{\rm s})$ for the transmitted
molecules. This is indeed observed in our experiment, and the
fringe contrast $V_\lambda=(S_{max}-S_{min})/(S_{max}+S_{min})$
serves to characterize the interference pattern.

The quantum origin of the observed signal is confirmed by the
characteristic dependence of its contrast $V_\lambda$ on the
molecular wavelength. If there is a coherent evolution in the
interferometer the expected fringe signal is easily calculated,
using wave optics in paraxial approximation. Starting with an
incoherent beam one finds, after a coherent passage from the
first to the third grating \cite{Brezger2002a},

\begin{equation}
\label{eq:signal}
  S(x_{\rm s}) \propto
  \!\!\sum_{m\in\Z}
  \left({B^{(0)}_m}^*\right)^2 B^{(\lambda)}_{2m}
\exp\left(2\pi\rmi m \frac{x_{\rm s}}{d}\right)
\end{equation}
The coefficients $B_m$ are defined in terms of the Fourier
components $a_\ell$ of the transmission function
\eref{eq:gratingfunction} of each of the three equal gratings,
\begin{align}
\label{eq:defBlambda} B_m^{(\lambda)}&=\sum_{\ell\in\Z} a_\ell
a^*_{\ell-m}
\exp\left(\rmi\pi\frac{m^2-2\ell
m}{2}\frac{L}{\LT}\right) \CO \intertext{and} \label{eq:defBnull}
B_m^{(0)}&=\sum_{\ell\in\Z} a_\ell a^*_{\ell-m} \PO
\end{align}
The $B_m^{(\lambda)}$ describe the diffraction at the second
grating, while the $B_m^{(0)}$ belong to the first and third
grating serving to mask the molecular density. The Fourier
coefficients $a_\ell$ also include the effect of the attractive
interaction between molecule and grating in eikonal approximation
\cite{Brezger2002a,Brezger2003a}. The masking by G$_1$ and G$_3$
is essentially wavelength independent. But due to the finite
flight time of the molecules through the grating slits the van der
Waals force introduces a certain wavelength dependence also in
the $B_m^{(0)}$.

Before taking the experimental signal as evidence for quantum
interference one must note that a certain fringe contrast could
also be explained by classical mechanics due to a shadow effect.
The classical expectation can be calculated by propagating the
classical phase space density of an uncollimated particle stream
through the interferometer subjected to the same forces and
approximations as in the quantum case. For ideal gratings the
resulting expression is given by Eq.~\eref{eq:signal} after
simply replacing the $B_{2m}^{(\lambda)}$ by $B_{2m}^{(0)}$
\cite{Hornbergerinprep}. In the presence of van der Waals
interactions  the deflection  tends to be underestimated in the
classical analog of the eikonal approximation, so that our
classical calculation gives an upper limit for the classical
visibility. Hence, whenever the experimental contrast is
significantly greater than the classical value one has evidence
that quantum interference took place.

An even stronger proof is the characteristic wavelength
dependence of the fringe visibility. Varying the mean molecular
velocity corresponds to changing the mean molecular de Broglie
wavelength. This dependence is used to scan the Talbot length
$L_\lambda$ and to demonstrate the periodic wave nature of the
molecular Talbot Lau effect in Fig.~\ref{fig:2}.

The experimental data are shown as full circles and are generally
well represented by the quantum theoretical calculation (solid line).
Both clearly show the expected recurrence of the visibility with
$\lambda$. The classical expectation (dash-dotted line)
completely fails to reproduce the observed effects. Note that the
visibility peaks are neither equally high nor symmetric around
their maxima. Also, the peaks do not occur exactly at the Talbot
length (indicated by arrows in Fig.~\ref{fig:2}). Instead the
maxima are shifted to shorter wavelengths and the peaks have a
broad shoulder towards the longer wavelengths. These deviations
from the simple optical Talbot Lau effect appearing in the
experiment are reproduced in the quantum calculation once we take
into account the retarded van der Waals
interaction~\cite{Casimir1948a} between the polarizable molecules
and the gold bars of the gratings. This interaction reduces the
effective slit width. At fixed grating distance this corresponds
to a shift of the maxima towards smaller wavelengths.

Fig.~\ref{fig:2} shows two theoretical predictions which
represent the experimental data. The dashed line assumes a
grating with an open fraction, i.e. a ratio of opening to grating
constant, of 0.48 as originally specified when the gratings were
purchased and mounted about two years ago. The solid line assumes
and open fraction of 0.45. A possible explanation for the
apparent shrinking of the openings might be a deposited layer of
fullerenes which are also visible to the unaided eye, at least on
the first grating. But also tiny mechanical grating deformations
might be a reason.

We observe a notable difference between theory and observation in
the peak height at a wavelength around 5\,pm, corresponding to
molecules with a mean velocity of around 100\,m/s. We have
evidence for the hypothesis that the experimental reduction of
the interference contrast is mainly due to remaining vibrations
of the setup with oscillation amplitudes of a few ten nanometers.
Further investigations of this effect are currently under way.

\section{Collisional decoherence: A quantum system interacting
with the environment} \label{sec:2DecoherenceTheory}

We now introduce a controlled source of decoherence by filling
the vacuum chamber with various gases at low pressure
($p=0.05\ldots 2.5\times 10^{-6}\,$mbar) at room temperature. Each
collision between a fullerene molecule and a gas particle
entangles their motional states. Hence, the effect of a single
collision on the molecular center-of-mass state is obtained by
tracing over the state of the scattered molecule. One can safely
assume that the mass of the fullerene molecule is much greater
than the mass $\mg$ of the gas particle. We then find that the
density operator, $\rho_0(\mathbf{r},\mathbf{r}')$, describing
the quantum state of the fullerene molecules, changes simply by a
multiplicative factor,
\begin{equation}
\label{eq:rho} \rho(\mathbf{r},\mathbf{r}')=\rho_0(\mathbf{r},
\mathbf{r}')\, \eta(|\mathbf{r}-\mathbf{r}'|)\PO
\end{equation}
This factor $\eta$ may be called the decoherence function since it
describes the effective loss of coherence in the fullerene state.
For elastic scattering with an isotropic potential and the gas
initially in a thermal state it reads  \cite{Hornberger2003b}

\begin{eqnarray}
\label{eq:eta} \eta(\Deltar)&=& \int_0^{\infty} \!\rmd \vg
\;\frac{g(\vg )}{\sigma(\vg )} \int\rmd\Omega
\left|f\big(\cos(\theta)\big)\right|^2 \nnn && \times
\sinc\!\Big(\sin\!\Big(\frac{\theta}{2}\Big)\frac{2 \mg \vg
\Deltar}{\hbar} \Big) \PO
\end{eqnarray}
This expression involves an integration over the thermal
distribution $g(\vg)$ of the gas velocities and an integral over
the scattering angle $\Omega=(\theta,\phi)$. In the argument of
the sinc function one finds the distance of the considered points
times the momentum change in units of $\hbar$. Hence, the sinc
function suppresses the integrand whenever the change in the
state of the gas particle during a collision is able to resolve
the distance $\delta r$. This leads to a reduction of the
corresponding off-diagonal elements in \eref{eq:rho} when the gas
particle transmits (partial) position information about the
molecule  to the environment.  For small distances the sinc
approaches unity so that the angular integral yields the total
scattering cross section $\sigma(\vg)$. Hence, for $\delta r \to
0$ the decoherence function approaches unity as required from the
conservation of the trace in \eref{eq:rho}.

By formulating the molecular evolution through the interferometer
in the Wigner representation one finds that the effect of
collisional decoherence can be treated analytically
\cite{Hornbergerinprep}. It is completely described by a
modification of the coefficients \eref{eq:defBlambda}.
To obtain the interference signal in the presence of a gas
 the $B_{2m}^{(\lambda)}$ in \eref{eq:signal} must be replaced by
\begin{align}
\label{eq:Bdeco}
B_{2m}^{(\lambda)}
\exp\!\Big(-n\sigma_{\rm eff} \int_0^{2L}
\Big[1-\eta\Big({m}\frac{L-|z-L|}{\LT}d \Big)\Big]\rmd z\Big)
\PO
\end{align}
Here  $n\sigma_{\rm eff}$ is the number density of gas particles
times the effective total cross section defined below in
Eq.~\eref{eq:sigmaeff1}, describing the number of collisions per
unit length. The integral in the exponent covers the various
positions in the interferometer where a collision may occur. As
discussed above we have $\eta(0)=1$, and the function decreases
to zero for increasing arguments. It follows that the $m=0$
component, related to the mean flux through the interferometer,
is not affected by decoherence. The other components of the
interference signal are most sensitive to collisions occurring
close to the second grating, at $z=L$, where the path separation
is greatest. Indeed, if the Talbot criterion is met, $L=\ell\LT$,
$\ell\in\Natural$,
 the distances entering the decoherence function are integer
multiples of the grating period $d$. For the other $z$ positions
the sensitivity decreases according to the path separation and,
as one expects, a collision event will not contribute to decoherence
directly at the first or at the third grating, at $z=0,\,2L$.

One can show that the general formula \eref{eq:Bdeco} is
equivalent\footnote{As shown in \cite{Hornberger2003b} the
original master equation by Gallis and Fleming \cite{Gallis1990a}
predicts a localization rate that is too large by a factor of
$2\pi$. It would yield an additional $2\pi$ in the exponent of
Eq.~\eref{eq:Bdeco}. The experimental results discussed in
Section 4 are sensitive to this factor (and rule it out).} to the
solution in paraxial approximation of the master equation for the
decoherence of a massive particle by interacting with a gas
\cite{Gallis1990a,Hornberger2003b}.
The present form
\eref{eq:Bdeco} has the advantage that it separates the rate of
the decoherence events, the factor  $n\sigma_{\rm eff}$, from
their effect described by the integrand in \eref{eq:Bdeco}. This
is particularly useful if the two processes should be  treated
with different degrees of accuracy, as is the case in the present
experiment.

At the prevailing (room) temperature of our experiment each
collision with a gas particle is so strong that it serves to
localize the fullerene to the scale of a few nanometers, which is
small compared to the typical path separation of 1\,$\mu$m.
Therefore one can safely approximate the integral in
\eref{eq:Bdeco} by $2L$ if $m\neq 0$. On the other hand, the
effective scattering cross section must be evaluated with care,
since it must account for the longitudinal velocity $v_{\rm
m}\mathbf{e}_z$ of the fullerene and for the thermal distribution
$\mu(\mathbf{v}_{\rm g})$ of the  gas particle velocities. The
general expression reads
\begin{align}
\label{eq:sigmaeff1}
\sigma_{\rm eff}(\vm) =& \int  \mu(\mathbf{v}_{\rm g})
\sigma(|v_{\rm m}\mathbf{e}_z-\mathbf{v}_{\rm g}|) \frac{|v_{\rm
m}\mathbf{e}_z-\mathbf{v}_{\rm g}|}{v_{\rm m}}
\,\rmd\mathbf{v}_{\rm g}
\PO
\end{align}
In our experiment the interaction potential is well described  by
the isotropic London dispersion force (van der Waals force
between polarizable molecules). The corresponding potential
$U(r)=-C_6/r^6$ has a single parameter $C_6$ that can be found in
\cite{Hornberger2003a} for a number of gases. The cross section
$\sigma(v)$ for a fixed relative velocity follows from a
semiclassical calculation \cite{Maitland1981a} and the remaining
integration in \eref{eq:sigmaeff1} can be performed asymptotically.
One finds
\begin{align}
\sigma_{\rm eff}(\vm) =
\frac{2(3\pi^6 C_6/8\hbar)^{2/5}}{\pi^\oh\Gamma(2/5)\sin(\pi/5)}
\,\frac{\vgt^{3/5}}{\vm}\, G\!\left(\frac{\vm}{\vgt }\right)
\label{eq:sigmaeff2}
\end{align}
with
$\vgt
= ({2\kB T/\mg})^\oh$ the most probable velocity in the gas and
\begin{align}
G(u)=
\Gammafunction(9/5) (1-u^2)
+\frac{2}{3}\Gammafunction({14}/{5})u^2+\Or(u^4)
\PO
\end{align}
This effective cross section exceeds
the geometric one by two orders of magnitude at the
velocities of our experiment ($\vm=80\ldots 240\, {\rm
m/sec}$).

A further simplification comes from the fact that for our
experimental setup the visibility of the interference signal
\eref{eq:signal} is essentially determined by the $m=0$ and
$m=\pm1$ Fourier components only. The expected reduction of the
contrast is therefore easily evaluated in terms of the
coefficients \eref{eq:Bdeco} and \eref{eq:defBnull}. Using $p=n
k_{\rm B}T$ one finds
\begin{eqnarray}
\label{eq:Vdeco} V_\lambda(p)&=& \frac{2
|B^{(\lambda)}_{2}|(B^{(0)}_1)^2}{(B^{(0)}_0)^3}
\exp\left(-\frac{2L\sigma_{\rm eff}}{\kB T}p\right) \nnn &=&
V_\lambda(0)\exp(-p/\pv)
\PO
\end{eqnarray}
Hence, we expect an exponential decrease of the visibility as a
function of the gas pressure $p$. This is the expected
experimental signature of collisional decoherence. It should not
be confused with Beer's law for absorption which predicts an
exponential decrease of the mean signal at \emph{constant}
visibility.

\section{Experimental decoherence: the pressurized interferometer}
\begin{figure}[b]
  \centering
  \includegraphics[width=\columnwidth]{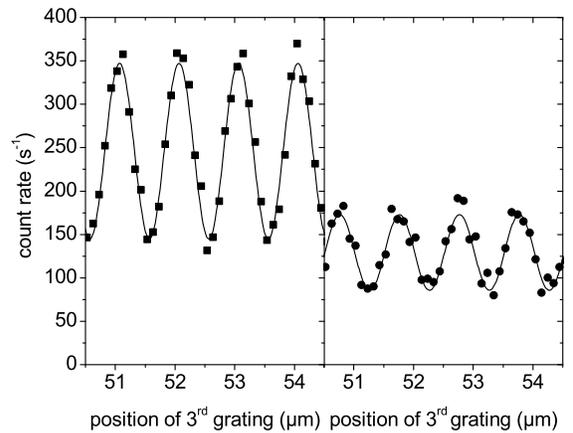}
\caption{Left: C$_{70}$ interference fringes at a pressure of
$3\times 10^{-8}$\,mbar (residual background gases) shown as full
circles. Right: the same signal in the presence of argon gas, at
a pressure of $5\times 10^{-7}$ mbar. The lines are fits of a
sine function. The mean velocity of the fullerene molecules was
189\,m/sec.} \label{fig:3}
\end{figure}

A first experimental indication of collisional localization is
presented in Fig.~\ref{fig:3}. It shows the change in the
interference pattern of C$_{70}$ if  a small amount of argon gas
is added to the vacuum chamber. We observe a significant
reduction in visibility from 42\% to 34\% if the pressure in the
chamber is increased from $3 \times 10^{-8}$\,mbar to $5 \times
10^{-7}$\,mbar. The horizontal shift between the two curves is
not significant since it can be explained by thermal drifts of
the setup between the two recordings. In contrast to that, the
 values of the visibilities \emph{are} significant. We tested
that they were reproducible within  $\pm$\,2 percent on different
days over several weeks.

\begin{figure}
  \centering
  \includegraphics[width=\columnwidth]{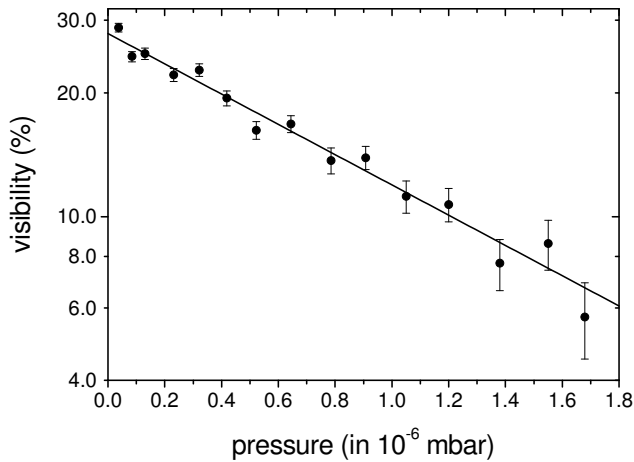}
\caption{Visibility of the  C$_{70}$ fringes as a function of
the  argon pressure at room temperature. The experimental data are
given by full circles and the theoretical prediction
\eref{eq:Vdeco} gives the slope of the solid line. The data
follow very nicely the expected exponential decay proving the
occurrence of collisional decoherence. This experiment was done
with L=22\,cm} \label{fig:4}
\end{figure}

We then record the visibility for a series of interferograms at
different gas pressures. A typical result is given in
Fig.~\ref{fig:4}, which shows the pressure dependence of the
interference visibility in the presence of thermal argon gas.
As expected from Eq.~(\ref{eq:Vdeco}) we observe an exponential
decay. Given the high initial contrast this is clear evidence
for the occurrence of collisional
decoherence.

The good quantitative agreement with decoherence theory is obtained
after taking into account a modification that is related to the
particular method of velocity selection used in the experiment.
As discussed in Sect.~\ref{sec:1Setup} we employ a gravitational
velocity selection scheme by restricting the molecular beam to a
free-flight parabola. If the apparatus is filled with a gas
this velocity selection gets disturbed by
collisions \emph{outside} of the interferometer.
After a collision each molecule gets slightly deflected so that now fullerenes
with a `wrong' velocity may fit through the setup. At the same
time the molecule detector has only a finite size so that some of
the molecules will pass it undetected after a collision.
One has to take into account the small modification of the expected decay
of visibility due to these effects. We do this by solving the \emph{classical} phase space dynamics, i.e., the Boltzmann
equation, effectively by a  Monte Carlo method. The scattering
angles are determined from the (diffraction limited)
differential cross sections. Semiclassical
expressions for the latter can be found in \cite{Helbing1964a}.
Our predictions for the visibility are obtained by weighting
\eref{eq:Vdeco} with the classical velocity distribution in the
detector -- which corresponds to an averaging over a distribution
of de Broglie wavelengths. Also the reduction of the mean count
rate found in Fig.~\ref{fig:2} is well reproduced by our
calculation.

\begin{table*}[bt]
\caption{Decoherence pressures, i.e. pressures for $V_\lambda(p)/V_\lambda(0)= e^{-1}$
for various collision gases as found in the short interferometer
with L=22\,cm, at a mean molecular speed of $\vm=117\,{\rm
m/sec}$. Note the very weak dependence on the mass of the
colliding partner.  All pressures given in units of
$10^{-7}$\,mbar.} \label{tab:1} \center
\begin{tabular}{rcccccccccccc}
\hline\noalign{\smallskip}
Atom & H$_2$ & D$_2$& He & CH$_4$& Ne & N$_2$ & Air & Ar & CO$_2$ & Kr & Xe & SF$_6$\\
\noalign{\smallskip}\hline\noalign{\smallskip}
mass/amu& 2 & 4 & 4 & 16 &   20.2& 28& 28.8& 39.9& 44&83.8&131.3&146 \\
$p_{0}$ (theo.)& 7.3 &  9.2 & 13.8 &  7.9&   16.0 &11.3&11.3&11.8&N.A.&12.4&11.5&N.A.\\
$p_{0}$ (exp.)& 4.6 &  8.0 & 10.7 &  8.1&   13.2 &11.5&10.5&10.8&8.9&12.9&10.6&11.3\\
$\Delta p_{0}$ (exp.)& 0.7&  1.2 & 1.6 &  1.2&   2.0& 1.7 & 1.6  & 1.6 & 1.3& 1.9&1.6& 1.7\\
\noalign{\smallskip}\hline
\end{tabular}
\end{table*}

The loss of coherence with increasing pressure is conveniently
described by the 'decoherence pressure' $p_0$ defined in
\eref{eq:Vdeco}. Table~\ref{tab:1} compares the measured values
of $p_0$ to the theoretical predictions for a number of gases.
One observes satisfactory agreement over the whole mass range
which covers two orders of magnitude. The experimental error
is mainly due to the uncertainty in the
pressure measurement, which is about 15\%.

The most remarkable feature of the  results reported in
Tab.~\ref{tab:1} is the very weak dependence of the decoherence
pressure on the type of gas used. This can be explained by
assuming that the polarizability of the gas particle is
proportional to its mass $\mg$. Then also  $C_6$ is proportional
to $\mg$, and one observes from \eref{eq:sigmaeff2} that the mass
dependencies of the interaction constant and of the most probable
gas velocity $\vgt$ almost cancel  out leaving $\sigma_{\rm
eff}\sim\mg^{1/10}$. The observed variations in Tab.~\ref{tab:1}
are due to deviations from the assumed proportionality and
reflect the specific electronic structures of the gas particle.

\begin{figure} [htb]
  \centering
  \includegraphics[width=\columnwidth]{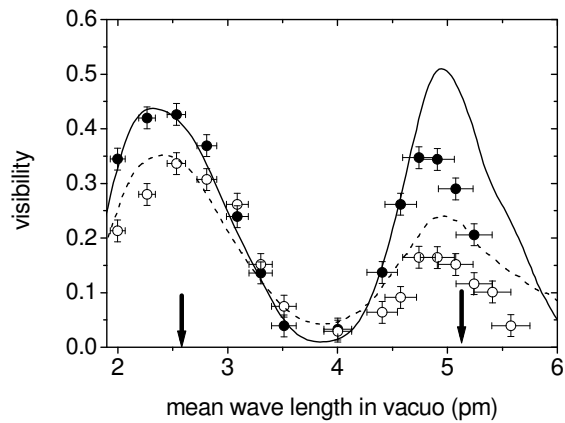}
\caption{Pressure dependence of the visibility in the stretched
Talbot Lau interferometer (height delimiter at $150\,\mu$m). The
main feature in this plot is the strong dependence of the
interference visibility on the molecular wavelength, i.e. the
molecular velocity. As expected, the slow molecules are much
stronger affected by collisions than the fast ones. The
experimental and theoretical curve for a background pressure of
$p=3\times 10^{-8}$\,mbar are given by the full circles and a
solid line. The hollow circles and the dashed line represent the
data for argon at $p=5\times 10^{-7}$\,mbar. The overall good
agreement between theory and experiment deteriorates markedly in
the regime of long wavelengths, i.e. small velocities, where
residual vibrations of the interferometer are expected to be
relevant.} \label{fig:5}
\end{figure}

It should be noted that the best contrast for all experiments
contributing to Tab.~\ref{tab:1} was systematically smaller than
that of Fig.~\ref{fig:2}. Although the whole experiment was
mounted on top of an optical table with active pneumatic
vibration isolation we were able to identify tiny vibrations of
the interferometer, induced by the water flow in the laser
cooling system, as being the cause of a reduced contrast.  The
visibility was increased by about 10\% when the laser was set on
rubber feet. This simple remedy was applied for the experiments
in the extended interferometer (L=38\,cm) of
Figs.\,\ref{fig:1},\ref{fig:2},\ref{fig:3},\ref{fig:5} but not
yet for the experiments of Fig.~\ref{fig:4} and Table~\ref{tab:1}
(L=22\,cm).

However, it is  important to note that the influence of external
vibrations and the effect of collisional decoherence are
independent of each other, and their total effect can therefore be
obtained by multiplying their respective contributions to the
reduction in visibility. This assumption was experimentally
verified for several gases: a varying initial (low-pressure)
contrast always led to the same slope in the
visibility-vs-pressure curve, i.e., to the same decoherence
pressure. The validity of Table~\ref{tab:1} is therefore not
compromised by potential mechanical perturbations.

However, the total time of flight plays an important role for the
absolute value of the fringe contrast. Clearly, for slow
molecules the interaction time with the gratings is longer, and
vibrations will be more detrimental. Also, the effective cross
section \eref{eq:sigmaeff2} increases for decreasing molecular
velocities. In Fig.~\ref{fig:5} we observe the increasing
influence of collisions in an interferometer filled with argon,
for various fullerene velocities, i.e. various interaction times.
In contrast to the experiment of Fig.~\ref{fig:2}, the central
height delimiter was set to 150\,$\mu$m (instead of 50 $\mu$m).
This increased the flux but it also reduced the overall contrast
by a few percent, both due to the increased sensitivity to
imperfections in the grating alignment and due to an increased
width of the velocity distribution.

Fig.~\ref{fig:5} shows the experimental visibility curves for
$p_{\rm low}=3\times 10^{-8}$\,mbar (full circles) and $p_{\rm
high}=5\times 10^{-7}$\,mbar (hollow circles) and compares them to
the quantum calculation (solid and dashed line, respectively)
with the same model parameters as already used for
Fig.~\ref{fig:2}. The remarks of the discussion of
Fig.~\ref{fig:2} apply also here. This holds for the reduction of
the visibility at long wavelengths due to vibrations, the shift
of the maxima with respect to the Talbot length and the
asymmetric line shapes --- caused by the molecule-wall
interaction. The new feature  in this graph is the contrast
reduction due to the scattering events in the Argon atmosphere
and their dependence on the interaction time. For wavelengths in
the 2.5\,pm regime ($v\sim 200\,$m/s) the pressure increase leads
to  $V_\lambda(p_{\rm high})/V_\lambda(p_{\rm low})=0.8$ whereas
for wavelengths in the 5\,pm regime ($v\sim 100\,$m/s) the
increased pressure results in $V_\lambda(p_{\rm
high})/V_\lambda(p_{\rm low})=0.5\,$. These ratios are identical
for   theory and experiment, both for the slow and for the fast
molecules. However, the {\em absolute} values still differ at long
wavelengths. Again this shows that the different decoherence
mechanisms (collisions and vibrations) are independent of each
other and their effects can be considered as separate
multiplicative factors to the visibility.

\section{Conclusion}
Our experiments have demonstrated the periodic nature of the
Talbot Lau effect in a molecule interferometer by observing a
visibility recurrence in the elongated setup (L=38\,cm). The
increased grating separation permitted a detailed quantitative
study of decoherence due to collisions with the background gas.
Our experiments show that decoherence by scattering of small
particles, which is ubiquitous in our macroscopic world, can be
understood and well controlled under high vacuum conditions.
Based on the good agreement which we found in comparing our
experiments with our numerical simulations we will now estimate
the residual gas pressures required to observe the quantum nature
of much larger objects.

To be specific, we consider a set of proteins of increasing size
up to the mass of a rhinovirus, interacting with molecular
nitrogen (300\,K, polarizability $\alpha/\Angstrom^3=4\pi\epsilon_0\times
1.75$). Since the static polarizability of large
hydrocarbons is closely proportional to their mass $M$, i.e.,
$\alpha/\Angstrom^3=4\pi\epsilon_0\times 0.123\,M/{\rm amu}$, we can use
the Slater-Kirkwood approximation \cite{Bernstein1979a}.

For the observation of interference with supermassive molecules
one needs low velocities in order to get de Broglie wavelengths
larger than 100\,fm. For particles in the mass range of
10$^5$\,amu this requires velocities of the order of
$\vm=10\,$m/s. Although this is a rather demanding requirement it
seems not impossible to develop appropriate sources in the future.
Moreover, a realistic earth-bound interferometer would be limited
to a Talbot length of $L\sim 1\,$m.

Based on these assumptions we extrapolate in Table~\ref{tab:2}
the decoherence pressures for insulin, green fluorescent protein,
hemoglobin, ferritin and a human rhinovirus. It turns out that
the vacuum conditions for quantum interference of these objects
can be provided using commercially available technology.

\begin{table*}
\caption{Estimated decoherence pressures of candidates for matter
wave interferometry in a stretched Talbot interferometer
($L=1$\,m, $\vm=10\,$m/sec) in the presence of air. The effective
cross sections are based on reasonable estimates for the
polarizability and the effective number of valence electrons.}
\label{tab:2}
\begin{tabular}{cccccccc}
\hline\noalign{\smallskip}
object & C$_{70}$ & insulin & GFP$^a$& hemoglobin& ferritin &virus$^b$\\
\noalign{\smallskip}\hline\noalign{\smallskip}
mass (amu)&   840  & 5730   &  $2.7\times 10^4$ & $6.4\times 10^4$ & $4.8\times 10^5$& $8\times 10^6$\\
min/max extension (nm) & 1 & 3 & 3/4 & 5/7 & 10& 30 \\
estim. $\sigma_{\rm eff}$(nm$^2$)   &730  & 1900 &  3700&  5200 & $1.1\times 10^4$& $3.6\times 10^4$  \\
estim. $p_{0}$ (mbar)  &$3\times 10^{-8}$& $1\times 10^{-8}$  & $6\times 10^{-9}$&  $4\times 10^{-9}$ & $2\times 10^{-9}$ & $6\times 10^{-10}$ \\
\noalign{\smallskip}\hline
\end{tabular}\\
{\footnotesize a) green fluorescent protein\\
b) rhinovirus HRV2 S150}
\end{table*}

\section*{Acknowledgements}
We are grateful for useful discussions on decoherence with John
Sipe, Toronto, and acknowledge experimental help in the setup of
the experiment by Stefan Uttenthaler. This work was supported by
the Austrian FWF in the programs START Y177 and SFB 1505. BB has
been supported by a EU Marie Curie fellowship (No.\
HPMF-CT-2000-00797), and KH by the DFG Emmy Noether program. We
acknowledge contributions by the EU under contract
HPRN-CT-2002-00309.


\begin{thebibliography}{10}

\bibitem{Berman1997a}
P.~R. Berman, Ed.: {\em Atom Interferometry} \newblock(Acad.
Press, New York, 1997)

\bibitem{Arndt1999a}
M.~Arndt, O.~Nairz, J.~Voss-Andreae, C.~Keller, G.~V. der Zouw,
and
  A.~Zeilinger: Nature, {\bf 401}, 680 (1999). 

\bibitem{Arndt2002a}
M.~Arndt, O.~Nairz, and A.~Zeilinger in {\em Quantum
{[Un]Speakables}},
  R.~Bertlmann and A.~Zeilinger, Eds.;
\newblock (Springer, Berlin, 2002;
\newblock pages 333--351)

\bibitem{Joos1985a}
E.~Joos and H.~D. Zeh: Z. Phys. B., {\bf 59}, 223 (1985).

\bibitem{Zurek1991a}
W.~H. Zurek: Phys. Today, {\bf 44}(10), 36  (1991).

\bibitem{Chapman1995b}
M.~S. Chapman, C.~R. Ekstrom, T.~D. Hammond, R.~A. Rubenstein,
J.~Schmiedmayer,
  S.~Wehinger, and D.~E. Pritchard: Phys. Rev. Lett., {\bf 74},
  4783 (1995). 

\bibitem{Hornberger2003a}
K.~Hornberger, S.~Uttenthaler, B.~Brezger, L.~Hackerm{\"u}ller,
M.~Arndt, and
  A.~Zeilinger: Phys. Rev. Lett., {\bf 90}, 160401 (2003).

\bibitem{Myatt2000a}
C.~J. Myatt, B.~E. King, Q.~A. Turchette, C.~A. Sackett,
D.~Kielpinski, W.~M.
  Itano, C.~Monroe, and D.~J. Wineland: Nature, {\bf 403}, 269 (2000).

\bibitem{Reynaud2002a}
S.~Reynaud, B.~Lamine, A.~Lambrecht, P.~M. Neto, and M.-T. Jaekel:
Int. J. Mod. Phys., {\bf A17}, 1003 (2002). 

\bibitem{Nairz2000a}
O.~Nairz, M.~Arndt, and A.~Zeilinger: J. Mod. Opt., {\bf 47}, 2811
  (2000). 

\bibitem{Schollkopf1994a}
W.~Sch{\"o}llkopf and J.~P. Toennies: Science, {\bf 266}, 1345
  (1994). 

\bibitem{Borde1994a}
C.~Bord{\'e}, N.~Courtier, F.~D. Burck, A.~Goncharov, and
M.~Gorlicki:
  Phys. Lett. A, {\bf 188}, 187 (1994).

\bibitem{Lisdat2000a}
C.~Lisdat, M.~Frank, H.~Kn{\"o}ckel, M.-L. Almazor, and
E.~Tiemann: Eur.
  Phys. J. D, {\bf 12}, 235 (2000). 

\bibitem{Bruehl2002a}
R.~Br{\"u}hl, P.~Fouquet, R.~E. Grisenti, J.~P. Toennies, G.~C.
Hegerfeldt,
  T.~K{\"o}hler, M.~Stoll, and C.~Walter: Europhys. Lett., {\bf 59}, 357
  (2002). 

\bibitem{Clauser1992a}
J.~Clauser and M.~Reinsch: Applied Physics B, {\bf 54}, 380
(1992). 

\bibitem{Patorski1989a}
K.~Patorski in {\em Progress in Optics {XXVII}}, E.~Wolf, Ed.;
\newblock (Elsevier Science Publishers B. V., Amsterdam, 1989;
\newblock pages 2--108.)

\bibitem{Chapman1995c}
M.~S. Chapman, C.~R. Ekstrom, , T.~D. Hammond, J.~Schmiedmayer,
B.~E. Tannian,
  S.~Wehinger, and D.~E. Pritchard: Phys. Rev. A, {\bf 51}, R14 (1995).

\bibitem{Deng1999a}
L.~Deng, E.~W. Hagley, J.~Wen, M.~Trippenbach, Y.~Band, P.~S.
Julienne, J.~E.
  Simsarian, K.~Helmerson, S.~L. Rolston, and W.~D. Phillips: Nature, {\bf
  398}, 218 (1999).

\bibitem{Clauser1994a}
J.~F. Clauser and S.~Li: Phys. Rev. A, {\bf 49}, R2213 (1994).

\bibitem{Brezger2002a}
B.~Brezger, L.~Hackerm{\"u}ller, S.~Uttenthaler, J.~Petschinka,
M.~Arndt, and
  A.~Zeilinger: Phys. Rev. Lett., {\bf 88}, 100404 (2002). 

\bibitem{Brezger2003a}
B.~Brezger, M.~Arndt, and A.~Zeilinger: J. Opt. B, {\bf 5}, S82
(2002).

\bibitem{Hornbergerinprep}
K.~Hornberger et~al. in preparation.

\bibitem{Casimir1948a}
H.~B.~G. Casimir and D.~Polder:  Phys. Rev., {\bf 73}, 360 (1948).

\bibitem{Hornberger2003b}
K.~Hornberger and J.~E. Sipe: Phys. Rev. A, {\bf 68}, 012105
(2003).

\bibitem{Gallis1990a}
M.~R. Gallis and G.~N. Fleming: Phys. Rev. A, {\bf 42}, 38 (1990).

\bibitem{Maitland1981a}
G.~C. Maitland, M.~Rigby, E.~B. Smith, and W.~A. Wakeham, {\em
Intermolecular
  Forces - Their Origin and Determination} (Clarenden Press, Oxford,
  1981).

\bibitem{Helbing1964a}
R.~Helbing and H.~Pauly: Z. Phys., {\bf 179}(0), 16 (1964).

\bibitem{Bernstein1979a}
R.~B. Bernstein, {\em Atom-Molecule Collision Theory} (Plenum,
New York, 1979).


\end{thebibliography}
\end{document}